\documentclass[aps,prb,twocolumn,amsmath,amssymb,showpacs]{revtex4-1}
\usepackage[english]{babel}
\usepackage{color}
\usepackage{graphicx}
\usepackage{epsfig}
\usepackage{dcolumn}
\usepackage{bm}

\begin{document}
\title{Thermopower and dynamical Coulomb blockade in non-classical environments}
\author{M. Mecklenburg, B. Kubala, and J. Ankerhold}
\affiliation{Institute for Complex Quantum Systems and IQST, University of Ulm, 89069 Ulm, Germany}


\begin{abstract}
Charge and heat transfer through a nanoscale conductor is not only determined by the transmission properties of the electrons, but can also be strongly impacted by coupling to other degrees of freedom in the environment of the conductor.
Here, we analyze the influence of the electromagnetic environment on a simple, yet significant thermoelectric property, the thermopower, in the simple transport scenario of single-charge transfer across a tunnel junction. Considering both thermal and out-of-equilibrium steady-state environments, we find that the thermopower can be strongly affected by the environmental state and can, in turn, act as a sensitive probe of environmental properties.
\end{abstract}
\maketitle

\section{Introduction}
Thermoelectric properties of nano-scale solid state circuits have received considerable attention in the last years, both experimentally as well as theoretically\cite{Giazotto2006,Dubi2011}. Particular focus has been put on their applicability and optimization as heat engines\cite{Ruokola2012,Jordan2013,Bergenfeldt2014,Erdman2017,Rossello2017} and thermoelectric devices like thermometers, refrigerators, and heat to current converters\cite{Galperin2009,Sanchez2011a,Entin-Wohlman2015,Sanchez2017}. This in turn necessitates a deeper understanding of the nature of the transfer of charge, heat and energy through respective structures\cite{Meschke2006,Jezouin2013,Altimiras2014,Svilans2016,Sanchez2017,Dutta2017}. In parallel, these developments have triggered new designs and manipulation techniques of electronic devices on the nano-scale\cite{Sanchez2011a,Lee2013,Svilans2016,Kuperman2017,Sanchez2017}. Typical set-ups include apart from fermionic leads additional bosonic thermal reservoirs (e.g.\ circuit impedances, phonons, vibrons) that give rise to inelastic charge transfer processes and can be used to improve the efficiency or the control of corresponding  thermoelectric devices\cite{Leijnse2010,Entin-Wohlman2010,Entin-Wohlman2012,Entin-Wohlman2015} complementing ideas exploiting particulars of the internal level structure of molecular junctions or quantum dots.

\begin{figure}[h]
	\centering
	\includegraphics[width=8cm]{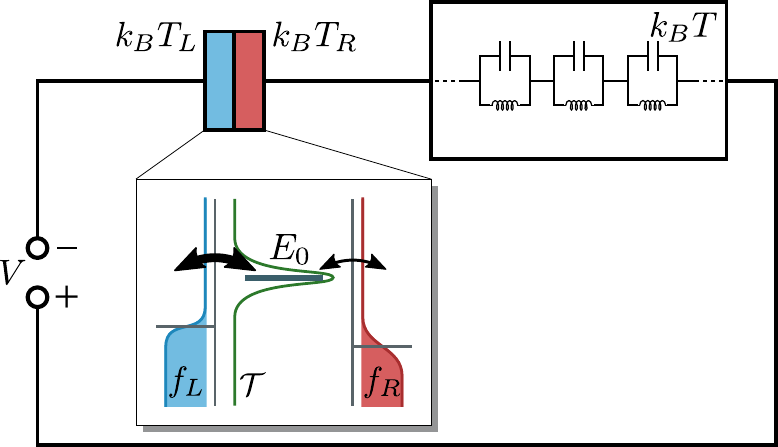}
	\caption{A tunnel junction with an applied temperature gradient
	is placed in series to an electromagnetic environment at temperature $k_BT$ and driven by a dc-voltage $V$.
	The electromagnetic environment can consist of arbitrary circuit elements and is modeled as a collection of $LC$-circuits.\\
	Energy dependent tunneling is modeled by a resonant level in the junction, which is  strongly tunnel-coupled to one side (inset).
	}
\label{fig:setup}
\end{figure}

In this work we study the impact of thermal and nonequilibrium steady-state environments on the {\em thermopower} of simple tunnel junctions in regimes, where single charge transfer prevails. These additional environments can be realized as specific circuit impedances placed in series with the junction such as ohmic resistors or $LC$-resonators, see Fig.~\ref{fig:setup}. They are conventionally assumed to reside in thermal equilibrium but one can also drive them to some well-defined steady state by external sources. That non-equilibrium state may even be a distinctively non-classical state, e.g., a single mode in a Fock-state. Such steady states have been intensively discussed in the context of reservoir engineering for quantum information processing devices and the tailoring of robust quantum states \cite{Mirrahimi2014,Souquet2014}. Also heat engines operated in presence of non-thermal reservoirs have been analyzed \cite{Rosnagel2014} but their influence on thermoelectric properties has been left untouched. In this respect, the thermopower plays a particular role as it is well-suited for sensitive measurements and allows to distinguish different transport regimes\cite{Koch2004,Segal2005,Scheibner2007,Reddy2007,Korol2016}.

Since in a first step we concentrate here on sequential charge transfer, the conventional description is based on the $P(E)$-theory \cite{Ingold1992} of dynamical Coulomb blockade which, however, is restricted to bosonic reservoirs in thermal equilibrium. Recently, an extension has been developed to capture also arbitrary steady state environments\cite{Souquet2014} of the type we are interested in the sequel. It turns out that this allows to arrive at a relatively transparent formulation for the thermopower of devices which are experimentally accessible. Single charge transfer can be achieved by circuit design but the validity of $P(E)$-theory and its extension also requires to operate devices in regimes, where, roughly speaking, a time scale separation exists between subsequent charge transfer events and the 'reset-time' of the environment (either relaxation time or time to restore the well-defined steady state).
Our findings are compared with standard results such as the Mott-relation which has been used previously to distinguish new features found in the thermopower from conventional ones \cite{Scheibner2005,Zotti2014,Ghahari2016}.

The outline of the paper is as follows. In Section \ref{sec:gen_phys} we introduce the general framework which is required to calculate the thermopower for given electromagnet environments with the $P(E)$-theory. As an illustrative example, in Section \ref{sec:ohmic}  purely ohmic environments in thermal equilibrium are considered. We then proceed in Section \ref{sec_non-classical} with reservoirs far from equilibrium such as Fock-states and squeezed states. Conclusion are drawn in Section \ref{sec:conclusion}.

\section{General theory}
\label{sec:gen_phys}
We start with a discussion of the main theoretical results which will serve as the basis for the more specific analysis later on. The general model is that of a tunnel contact which connects two metallic leads with chemical potentials $\mu_R$, $\mu_L$ and electronic temperatures $T_{L/R}=T_{\mathrm{el}}\mp \Delta T/2$, respectively, where $T_{\mathrm{el}}$ is the mean electron temperature with $|\Delta T|\ll T_{\mathrm{el}}$. Further details about the nature of the region between these leads will be given below.

\subsection{Thermoelectric transport in linear response}
In linear response the thermoelectric charge- and heat currents $I$ and $J$ are connected to the applied forces $eV=\mu_R - \mu_L$ and $\Delta T$, with $e=-|e|$ being the electron charge, via the linearized transport coefficients $\mathcal{L}_0$, $\mathcal{L}_1$ and $\mathcal{L}_2$, i.e.,
\begin{align}
\begin{pmatrix}
I \\ J
\end{pmatrix}
=
\begin{pmatrix}	
\mathcal{L}_0 & \mathcal{L}_1 \\
\mathcal{L}_1 & \mathcal{L}_2
\end{pmatrix}
\begin{pmatrix}
V \\ \frac{\Delta T}{T_{\mathrm{el}}}
\end{pmatrix}\, .
\end{align}
Onsager relations require, that without magnetic fields the off-diagonal (linear) thermoelectric coefficients are identical. Here,  the electrical conductance $G = \mathcal{L}_0$ and $G_T = T_{\mathrm{el}} \cdot\mathcal{L}_1$ define the thermopower  (or Seebeck coefficient)
\begin{align}\label{eq:thermopower}
S=-\left.\frac{G_T}{G}\right|_{I=0}\, ,
\end{align} 
which is a measure for the thermoelectrical voltage induced by a temperature gradient $\Delta T$ across the tunnel contact. In terms of the electrical current $I(V, \Delta T)$ the coefficients $G$ and $G_T$ read
\begin{equation}
	G=\frac{\partial\, I(V,0)}{\partial V} \quad \text{and} \quad G_T= \frac{\partial\, I(0,\Delta T)}{\partial \Delta T}\, .
	\label{eq:conductances}
\end{equation}

In the simplest situation of elastic tunneling through a channel with transmission probability $\mathcal{T}(E)$, the charge current is
\begin{align}
I(V, \Delta T)= e\,  \int\limits_{-\infty}^{\infty} \mathrm{d}E \, \left[ f_L(E) - f_R(E) \right] \mathcal{T}(E)\, ,
\label{eq:current-el}
\end{align}
where $f_L$ ($f_R$) is the Fermi function of the left (right) lead. Here, these are given by
\begin{align}\label{eq:fermi}
f_{L/R}(E)= \left\{1+\exp\left[\frac{E\pm e V/2}{k_{\rm B} (T_{\mathrm{el}}\mp \Delta T/2)}\right]\right\}^{-1}\, 
\end{align}
with energies counted with respect to the mean chemical potential $(\mu_L+\mu_R)/2$.
The expression for the thermopower then follows immediately from $G=\mathcal{E}_0$ and $G_T=\mathcal{E}_1$ with
\begin{align}\label{eq:elasticgs}
\mathcal{E}_\alpha = \frac{e^{2-\alpha}}{T_{\mathrm{el}}^{\alpha}} \int_{-\infty}^\infty \mathrm{d}E\,  \mathcal{T}(E) f_0'(E) E^\alpha\ ,\ \alpha=0, 1
\end{align}
where  $f^{\prime}_0=\mathrm{d}f_0/\mathrm{d}E$ and $f_0$ refers to the Fermi function (\ref{eq:fermi}) at $V=\Delta T=0$. This also shows that the thermopower can be understood as a measure for the mean transferred energy per tunneling event, one of the fundamental indicators for the efficiency of heat transfer. 

Further simplifications arise in the limit of very low temperatures, where the derivative of the Fermi function becomes localized so that an Sommerfeld-expansion around the Fermi energy can be performed. Due to the asymmetry of $E f'(E)$, only odd powers of $E$ in the expansion of $\mathcal{T}(E)$ are relevant for $\mathcal{E}_1$. Accordingly, one finds with $\sigma_0 =(\pi^2/3) k_{\rm B}/e $ the known Mott-result \cite{Heikkilae2013}
\begin{align}\label{eq:s0}
S_0\equiv S_{T_{\mathrm{el}}\to 0}= - \sigma_0 k_{\rm B} T_{\mathrm{el}} \frac{\mathcal{T}'(0)}{\mathcal{T}(0)}\, .
\end{align}

In an actual electrical circuit and at low temperatures, the coupling of the electromagnetic modes of the surrounding circuitry  with the tunneling charges cannot be neglected. In fact, if the charging energy of the junction $E_C=e^2/2C$ sufficiently exceeds the thermal energy scale $k_{\rm B}T$, dynamical Coulomb blockade may occur if the tunneling resistance $R_T\gg R_Q$ with $R_Q=h/e^2$ being the resistance quantum. This has led to the so-called $P(E)$-theory, where $P(E)$ captures the probability of tunneling electrons to absorb/emit energy from surrounding electromagnetic degrees of freedom.

The goal of the present work is to extend this framework to analyze heat transfer processes in thermal and designed quantum environments. According to (\ref{eq:s0}), the asymmetry of the transmission in the neighborhood of the Fermi energy is essential for a finite thermopower as otherwise forward and backward charge flows compensate each other. As a minimal model we thus consider a tunnel junction, where an intermediate resonant level at energy $E_0$ couples strongly asymmetrically to the left and the right leads, i.e., $\Gamma_L\gg \Gamma_R$ with individual coupling rates $\Gamma_L$ and $\Gamma_R$, respectively (see also inset of Fig.~\ref{fig:setup}). As a consequence, states on the resonant level are in thermal equilibrium with the left lead and tunneling events from the resonant level to the right lead occur in an energy window around $E_0$. Due to the weak tunneling rate $\Gamma_R\ll 1/\nu_0$ ($\nu_0$ is the density of states), charge transfer may be influenced by inelastic processes according to the $P(E)$ treatment. As we will see, this in turn influences the asymmetry in the heat flow and allows one to tailor or even to tune thermoelectric properties of devices by reservoir engineering. The corresponding transmission in (\ref{eq:current-el}) is then given by
\begin{align}
\mathcal{T}(E)= 4 \frac{\Gamma_R}{\Gamma_L} \cdot \frac{(\Gamma_L/2)^2}{(\Gamma_L/2)^2+(E-E_0)^2} = 4 \frac{\Gamma_R}{\Gamma_L} \cdot \tau (E)
\end{align}
and, in a Green's function formalism, is proportional to the spectral function of the level in thermal equilibrium \cite{Flensberg2003}.

Now, following the lines of the $P(E)$ treatment \cite{Ingold1992}, the inelastic charge transfer rate for forward (left to right) processes reads
\begin{align}
\overrightarrow{\Gamma}(V, \Delta T)  = \frac{1}{e^2 R_T} \int\limits_{-\infty}^{\infty}\mathrm{d}E \mathrm{d}\bar{E} & \, f_L(E) \left[ 1- f_R(\bar{E}) \right]\nonumber\\
&\times P(E-\bar{E})\,  \tau(E)\, ,
\label{eq:Gamma_inel}
\end{align}
where $R_T= \hbar \Gamma_L/(8 \pi e^2 \Gamma_R )\gg R_Q$. The net current then follows from
\begin{equation}
I(V,\Delta T) = e \left[ \overrightarrow{\Gamma}(V,\Delta T) - \overleftarrow{\Gamma}(V,\Delta T) \right]\, ,
\label{eq:Iinel}
\end{equation}
where the backward rate is obtained from (\ref{eq:Gamma_inel}) by interchanging the indices for left and right lead and replacing $\tau(E)\to \tau(\bar{E})$. Of course, the known $P(E)$-result for a conventional tunnel junction is regained by putting $\tau(E)\to 1$. Further, for the case of coupling to a thermal reservoir with temperature $T\equiv T_{\mathrm{el}}$ and $\Delta T=0$, the rates $\overleftarrow{\Gamma}$ and $\overrightarrow{\Gamma}$ are related by detailed balance.

Now, together with (\ref{eq:conductances}) and (\ref{eq:thermopower}), one finds in generalization of (\ref{eq:elasticgs})

\begin{align}\label{eq:inepsi}
\mathcal{E}_\alpha^{\mathrm{inel}} = & \frac{e^{-\alpha}}{T_{\mathrm{el}}^\alpha R_T} \int\limits_{-\infty}^{\infty} \mathrm{d}E \mathrm{d}\bar{E} \ \frac{1}{2} \left[\tau(E) + \tau(\bar{E})\right] P(E -\bar{E}) \nonumber\\
& \times \left[ \bar{E}^\alpha f_0(E) f_0^{\prime}(\bar{E}) +{E}^\alpha f_0(-\bar{E})f_0^{\prime}(E) \right]
\end{align}
so that $S^{\mathrm{inel}}=-\mathcal{E}_1^{\mathrm{inel}}/\mathcal{E}_0^{\mathrm{inel}}$. Note that $\mathcal{E}_0^{\mathrm{inel}}$ is symmetric around the Fermi-level with respect to a transition from $E_0 \to -E_0$ whereas $\mathcal{E}_1^{\mathrm{inel}}$ is anti-symmetric.

Determining the linearized thermopower experimentally has traditionally been done by applying a temperature gradient and measuring the arising voltage in an open circuit. In our case, linearizing the current (\ref{eq:Iinel}) yields $I_{\mathrm{lin}}=I_0+I_V+I_{\Delta T}$ with an off-set current $I_0$ being independent of both $eV$ and $\Delta T$. 
This current is due to non-equilibrium fluctuations, which are rectified by the strongly asymmetric coupling of the leads to the transport level yielding  $\overrightarrow{\Gamma}(0, 0) \neq \overleftarrow{\Gamma}(0, 0)$ even for $V=\Delta T=0$.

The problem is circumvented by using the current heating technique, where an ac-current with frequency $f$ generates a temperature gradient which varies with frequency $2f$~ \cite{Svensson2012,Svensson2013,Thierschmann2013,Thierschmann2015}. 
Locking-in to this frequency provides access to the ac-voltages due to the applied temperature gradient.

\begin{widetext}
	Again in the low temperature range simplifications arise and a Sommerfeld-like expansion yields
	\begin{equation}
		S_0^{\mathrm{inel}}=-\sigma_0 k_BT_{\mathrm{el}} \frac{
			 \int \limits_{-\infty}^{\infty} \mathrm{d}E \, \left\lbrace 2 \tau^{\prime}(0) f_0(E) P(E) - \left[ \tau(E) - \tau(-E) \right] \cdot f_0(E)P^{\prime}(E)\right\rbrace
			 }{
			 \int\limits_{-\infty}^{\infty} \mathrm{d}E \, \left[ \tau(E) +2\tau(0) + \tau(-E) \right]f_0(E)P(E)
			 }\, .
		\label{eq:Sinel_SF_ST}
	\end{equation}	
\end{widetext}

\section{Purely ohmic impedance}
\label{sec:ohmic}
\noindent
As a first example, we consider a circuit consisting of a tunnel junction including a resonant level as explained above and in series with an ohmic impedance. The corresponding $P(E)$-function, which is calculated by Fourier-transforming a phase-correlation-function containing the system dependent total impedance $Z_t$ (see Ref.~\onlinecite{Ingold1992} for details), is of Gaussian form, i.e., 
\begin{equation}
P(E) = \frac{1}{\sqrt{4 \pi E_C k_BT}} \exp \mathopen{} \left[ - \frac{(E-E_C)^2}{4 E_C k_BT} \right] \mathclose{}\, ,
\end{equation}
where the charging energy $E_C$ controls the coupling strength between junction and environment. The temperature of this latter bosonic reservoir is denoted by $T$ and is assumed to be independently controllable from the leads mean electronic temperature $T_{\mathrm{el}}$. It is convenient to work with dimensionless units by scaling energies with $E_C$  so that $\theta_{\mathrm{el}} =k_{\rm B}T_{\mathrm{el}}/E_C$ is the dimensionless temperature of the leads in equilibrium, $\epsilon=E_0/E_C$ the dimensionless resonance energy, $\gamma=\Gamma_L/(2 E_C)$ the dimensionless width of the resonance, and $\theta=k_{\rm B} T/E_C$ the dimensionless temperature of the electromagnetic environment; it determines the width of the above $P(E)$ function, where $E>0$ corresponds to absorption from and $E<0$ emission to fermionic degrees of freedom. Of course, for $\theta\to 0$, only absorption is possible and $P(E)\to \delta(E-1)$ in dimensionless units.

\begin{figure}[t]
	\centering
	\includegraphics[width=8cm]{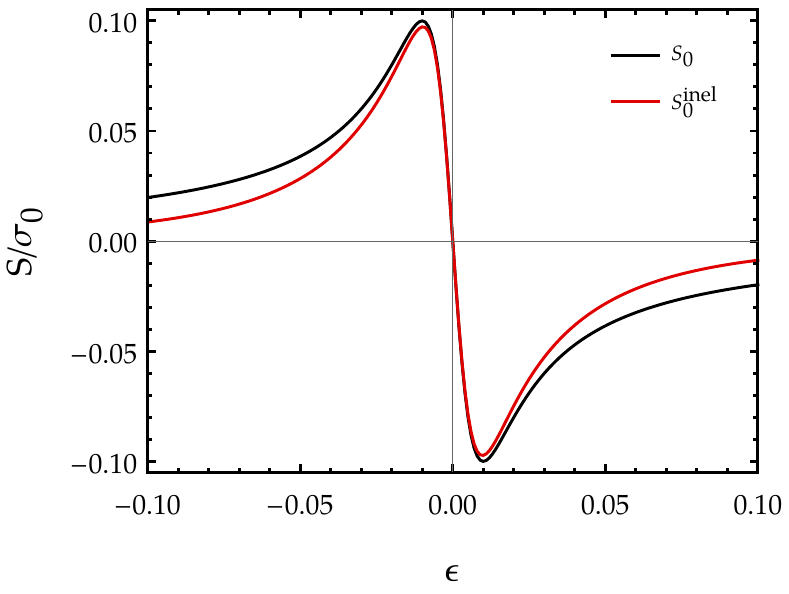}
	\caption{(Color online)
		Thermopower of a cold junction ($\gamma=0.01$, $\theta_{\mathrm{el}}=0.001$) coupled to a hot ohmic impedance ($\theta=1$) versus the level position $\epsilon$. The hot environment enables exchange processes over a wide range of energies. Nonetheless, the inelastic thermopower [red, Eq.~\eqref{eq:Sinel_SF_ST}] retains the Mott-form \eqref{eq:s0} of elastic transport  without such exchange processes (black). Units are scaled with $E_C$.
	}
	\label{fig:S_conf}
\end{figure}

Figure \ref{fig:S_conf} shows the thermopower in an environment consisting of an ohmic impedance at high temperature.
For small electronic temperatures the Sommerfeld result \eqref{eq:Sinel_SF_ST} (cf.~Fig.~\ref{fig:S_conf}) well approximates the full expression for $S^{\mathrm{inel}}$ resulting from \eqref{eq:inepsi} (not shown). The hot environment enables exchange processes over a wide range of energies. However, electrical and thermal conductance are affected in a similar manner, so that there is only a minor effect on their ratio. The thermopower then shows the familiar Mott-behaviour, $S\propto \tau'(0)/\tau(0)$, with extrema at $\epsilon\approx \pm \gamma$ with $S(|\epsilon|=\gamma)/\sigma_0\approx \theta_{\mathrm{el}}/\gamma$. The antisymmetric shape reflects the fact that below the Fermi energy more states are available in the hotter than in the colder reservoir, while above the opposite applies. By tuning the resonance energy (going from $-\epsilon$ to $\epsilon$) one thus sweeps the charge transfer from cold to hot to a transfer hot to cold.

\begin{figure}[t]
	\centering
	\includegraphics[width=8cm]{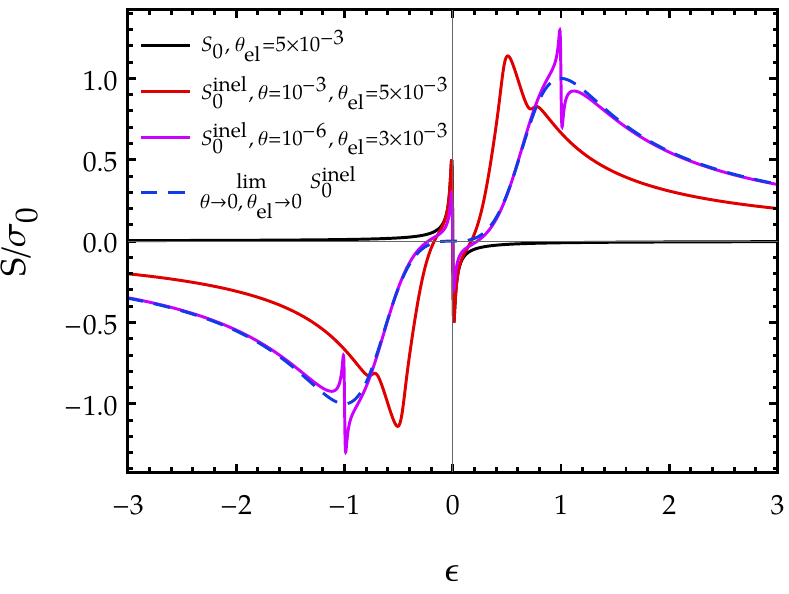}
	\caption{
		(Color online)
		In a low-temperature ohmic environment the charging energy $E_C$ appears as a new scale in the thermopower. In the $\theta_{\mathrm{el}}\ll 1$ and $\theta\to 0$ limit of Eq.~\eqref{eq:S_Theta_ST} a broad feature of width $E_C (=1)$ and additional copies of the sharp Mott-feature at $\epsilon \approx \pm 1$ combine (magenta curve). The crossover from the ballistic Mott result (black) for intermediate environmental temperatures is illustrated by the red curve and the limit of vanishing electronic and environmental temperature is shown (blue dashed line). Units are scaled with $E_C$ and $\gamma=0.01$.
	}
	\label{fig:S_ST_ohm_Tenv_dep}
\end{figure}

If the ohmic impedance is at very low temperature,  $\theta\ll 1$, so that in each tunneling process it will absorb exactly the charging energy, the pattern of the thermopower changes substantially  (see Fig.~\ref{fig:S_ST_ohm_Tenv_dep}) and reveals the impact of the newly relevant energy scale $E_C$. In general, two new features combine in the thermopower: a broad structure of width $E_C (=1)$, and two copies of the sharp Mott-like feature of width $\gamma$, which appear around $\epsilon\approx \pm 1$. These new features arise from the second term in the numerator of Eq.~(\ref{eq:Sinel_SF_ST}), with the copies of the Mott-features depending on the asymmetry in the transmission function $\tau(E)$ and the broad feature resulting from the asymmetry of the $P(E)$-function with respect to the Fermi-level. In the limit $\theta_{\mathrm{el}}\ll 1$ and $\theta\to 0$ the expression (\ref{eq:Sinel_SF_ST}) actually reduces to
\begin{align}
\lim_{\theta \to 0} \frac{S_0^{\mathrm{inel}}}{\sigma_0} =& -\theta_{\mathrm{el}} \frac{\tau^{\prime}(1) + 2\tau^{\prime}(0) + \tau^{\prime}(-1)}{\tau(1) + 2\tau(0) + \tau(-1)} \nonumber \\
&+ \frac{\tau(1) - \tau(-1)}{\tau(1) + 2\tau(0) + \tau(-1)}
\label{eq:S_Theta_ST}\;.
\end{align}
Here, the second term originates from the second contribution in (\ref{eq:Sinel_SF_ST}) and provides a broad antisymmetric profile on top of which the first contribution provides copies of standard Mott-like features at $\epsilon=0, \pm 1$. It is the second contribution which prevails in the limit $\theta_{\mathrm{el}}\to 0$ so that for $\gamma\ll 1$ one arrives at
\begin{equation}
\lim_{\theta, \theta_{\mathrm{el}} \to 0} \frac{S_0^{\mathrm{inel}}}{\sigma_0}	 \approx \left\lbrace\begin{matrix}
	\dfrac{2\epsilon^3}{(\epsilon^2 + 1)^2 + \epsilon^2 (\epsilon^2 + 1)} & \text{for } |\epsilon|\gg \gamma \\  & \\ 2\epsilon\gamma^2 + 2\epsilon^3 & \text{for } |\epsilon|\lesssim\gamma
	\end{matrix}\right. \, .
\end{equation}
Due to the strong hybridization between the left lead and the resonant level occupations of the level remain even for vanishingly small temperatures $\theta_{\mathrm{el}}$. These can emit energy into the environmental resonance,  when it is close to the level energy. While the coefficients $G$ and $G_T$ both vanish and there is no finite energy flow in this limit, their ratio yields a non-vanishing contribution to the thermopower. In Fig.~\ref{fig:S_ST_ohm_Tenv_dep} the cubic behaviour of the thermopower in a narrow region around the Fermi-level is clearly seen, while the broad profile with extrema at $\epsilon=\pm 1$ appears away from this domain.\\

Remarkably, in contrast to the ballistic case where a finite thermopower requires a non-zero electronic temperature, in the inelastic situation this is no longer true (blue dashed line in Fig.~\ref{fig:S_ST_ohm_Tenv_dep}). The newly appearing copies of the Mott-feature at $\epsilon \neq 0$ indicate the existence of new inelastic transport channels. Generally speaking, a strong signal appears in the thermopower, whenever a substantial part of transport is carried by a new channel, even though the overall probability of the considered process may be small. The thermopower (in the low-temperature Sommerfeld limit) is hence uniquely suited to indicate the presence of new transport channels. Accordingly, we will exploit it in the following to study the opening of new channels for inelastic transport by placing the junction in various non-classical environments. Note, that we will throughout consider the Sommerfeld approximation, where features of new channels are clearly distinguishable. Quantitative reliability of results can always be checked by full results based on Eq.~\eqref{eq:inepsi}.

\section{Non-classical environments}
\label{sec_non-classical}
The conventional $P(E)$-theory is based on the assumption of electromagnetic environments to reside in thermal equilibrium. Recently, it has been shown that the formulation can be generalized to environments in non-equilibrium stationary states, see Ref.~[\onlinecite{Souquet2014}] for details. In essence, one requires rare tunneling events to have only minor impact on the state of the reservoir such that its stationary state is re-established on a time scale sufficiently shorter than the time between subsequent tunneling events. Non-classical environments have received much attention recently in various context as discussed in the Introduction. Apart from their possible technological relevance, they may also open new ways to study system-reservoir correlations beyond standard settings. A typical example is a junction interacting with a cavity mode  with frequency $\Omega=1/\sqrt{LC}$ of an $LC$-circuit, where a desired cavity state, e.g.\ a Fock state or a coherent state, is maintained by appropriate external forces. The exact implementation of each such driving scheme then determines the internal time scales for the relaxation of the non-equilibrium to its steady state and the purity of the desired cavity state (cf. [23] considering finite quality cavities as a simple example of the latter). Here, it is convenient again to use dimensionless units, where now energies are measured in units of $\hbar\Omega$.

The $P(E)$-function describing energy exchange with the tunneling charge in this situation can then be formulated in terms of the probability distribution $P_0(E)$ to absorb energy by a ground state cavity and a quasi-probability distribution $P_{\mathrm{occ}}(E)$ attributed to the specific stationary cavity state. As a result one finds
\begin{align}
P_{\mathrm{tot}}(E) = & \int\mathrm{d}\bar{E} P_0(E-\bar{E}) P_{\mathrm{occ}}(\bar{E}) \nonumber \\
= & \sum\limits_{k=-\infty}^{\infty} p_{\mathrm{tot}}[k]\  \delta (E - k)\, .
\label{eq_P_tot}
\end{align}
To arrive at the second line, one exploits that
\begin{equation}
P_0(E) = \sum\limits_{k=0}^{\infty}e^{-\rho} \frac{\rho^k}{k!}\ \delta(E - k),
\label{eq_P0}
\end{equation}
with $\rho=E_C/\hbar\Omega$ for a cavity with frequency $\Omega$, playing the role of a coupling parameter between junction and cavity. The distribution $p_{\mathrm{tot}}[k]$ then describes the exchange of $k$ photons between tunneling charge and cavity given the distribution $P_{\mathrm{occ}}({E})$ (see Ref.~\onlinecite{Souquet2014} for details). We emphasize that $P_{\mathrm{occ}}({E})$ is not a true probability distribution as it can become negative. This particularly applies to stationary states which are far from being classical, e.g.\ Fock states. The above expression (\ref{eq_P_tot}) now allows one to analyze the influence of engineered quantum environments on the thermopower.

\begin{figure}[t]
	\centering
	\includegraphics[width=8cm]{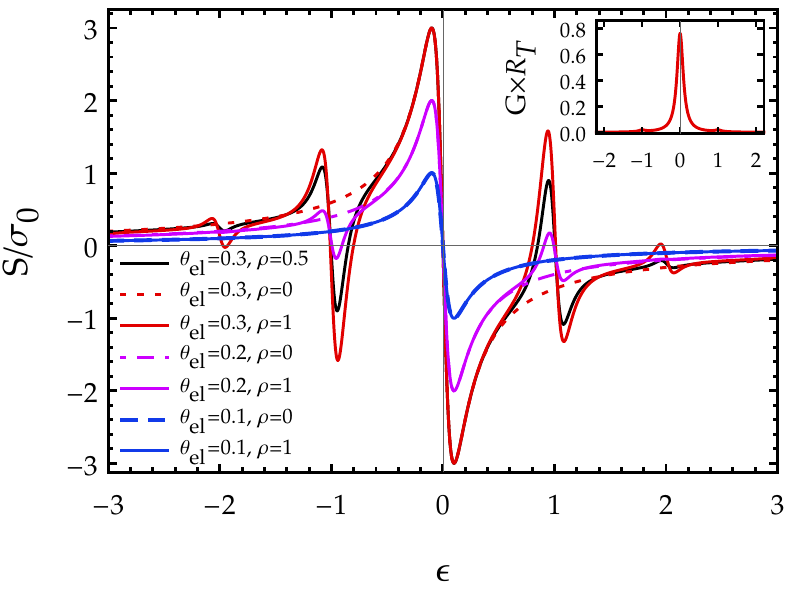}
	\caption{(Color online)
		Thermopower $S_0^{\mathrm{inel}}$ in Sommerfeld approximation (solid lines) within an electromagnetic environment consisting of a high-quality cavity in its ground state. Strong Mott-features in the thermopower at multiples of the cavity resonance, $\epsilon = \pm k$ $(k \in \mathbb{N})$, signal inelastic transport channels with $k$-photon emission into the cavity. Electronic temperature  $\theta_{\mathrm{el}}$ and coupling strength $\rho$ determine, whether transport through  an inelastic channel becomes dominant over the ballistic background (dashed lines). If so, there will be a large thermopower signal, even in cases where the total conductance (inset) may be small.  Units are scaled with $\hbar\Omega$ and $\gamma=0.1$.
	}
	\label{fig:S-ground_T-dep_simple}
\end{figure}

\subsection{Ground-state cavity}

We first consider a pure ground state preparation in the cavity, a situation which corresponds to $P_{\mathrm{occ}}({E})=\delta(E)$ in (\ref{eq_P_tot}) and can thus also be described within the conventional $P(E)$-treatment for a $LC$-impedance at zero environmental temperature $\theta$. The weights $p_{\mathrm{tot}}$ turn out as
\begin{equation}
p_{\mathrm{tot}}[k]=e^{-\rho} \frac{\rho^k}{k!}\, \Theta(k),
\end{equation}
with $\Theta(k)$ being the Heavyside step function.

The influence of the electronic temperature $\theta_{\mathrm{el}}$ as well as the coupling strength $\rho$ on the thermopower, Eq. (\ref{eq:Sinel_SF_ST}), is illustrated in Fig.~\ref{fig:S-ground_T-dep_simple}. With increasing electronic temperature (and finite $\rho$) new inelastic transport channels become accessible leading to additional peaks in the thermopower at multiples of the cavity's resonance energy. The comparison to the linear conductance (inset) demonstrates, how thermopower highlights a weak but dominant transport channel.

At lower temperatures ($\theta_{\mathrm{el}}=0.1$) deviations from the ballistic reference disappear: Inelastic processes are completely suppressed as the cavity can only absorb energy which the charge carriers are not able to provide. With increasing electronic temperature the Fermi distributions in the leads broaden and carriers with energy above the Fermi level can emit quanta into the cavity. The dependence on the junction-cavity coupling $\rho$ will be discussed in more detail below (Sec.~\ref{sec:rho-dependence}).

\begin{figure}[t]
	\centering
	\includegraphics[width=8cm]{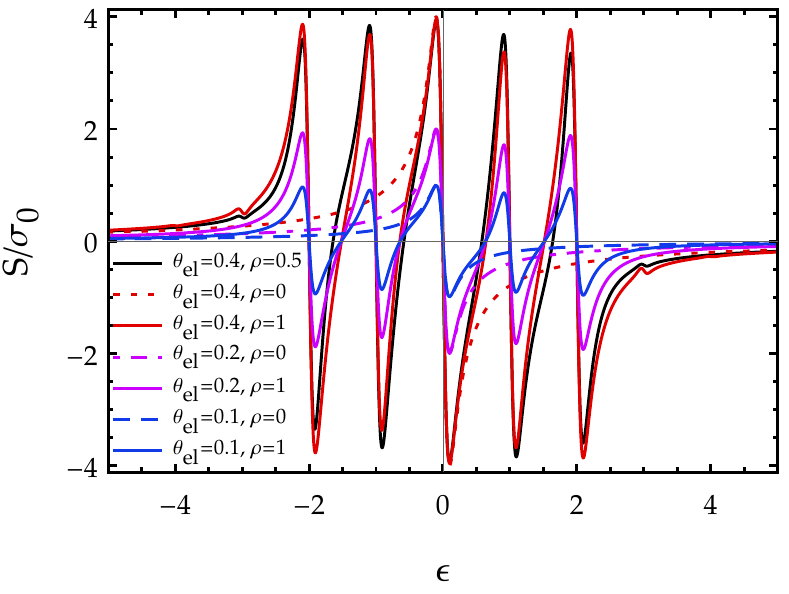}
	\caption{(Color online)
		Thermopower $S_0^{\mathrm{inel}}$ for a cavity in a Fock state $|n_0=2\rangle$. The non-classical environment opens new transport channels with strong Mott-features appearing at energies, $\epsilon = \pm k ~ (k \in \mathbb{N}, ~ k \le n_0)$. The low-temperature electronic system absorbs energy quanta from the environment lifting electrons from far below to the edge of the Fermi sea, where they yield a strong thermoelectric signal. Units are scaled by $\hbar\Omega$ and $\gamma=0.1$; dashed curves are ballistic results.
	}
	\label{fig_S-ST-fock_T-dep}
\end{figure}

\subsection{Fock-state cavity}
While an empty cavity can be considered as the zero-temperature limit of a classical (thermal) environment, we will now consider an environment of  manifest quantum nature, namely a cavity driven into a particular Fock-state $|n_0 \rangle$. Since there are now $n_0$ excitations available in the cavity, even at zero temperature $n_0$ quanta with energy $\hbar\Omega$ can be emitted. This allows transfer processes also below the Fermi-level. In particular, we consider $|n_0=2\rangle$, for which the weights in the non-equilibrium energy exchange function (\ref{eq_P_tot}) are given by

\begin{equation}\label{eq:p_tot_Fock}
p_{\mathrm{tot}}^{(2)}[k] = \frac{e^{-\rho} \rho^k \cdot 2!}{(k+2)!}\left[ L_2^{(k)}(\rho) \right]^2\,  \Theta(k+2)
\end{equation}
with $L_2$ being a generalized Laguerre polynomial.

We now find additional Mott-features above and below the Fermi level around the energies, $\epsilon = k$ $(k \in -n_0 \le \mathbb{N} \le n_0)$, see Fig.~\ref{fig_S-ST-fock_T-dep}. These peaks have absolute heights independent of the level position for $\rho=1$ (cf. Sec. Sec.~\ref{sec:rho-dependence} below for the $\rho-$dependence) and appear in doublets of opposite sign. In an actual set-up this allows to switch between a peak with $S>0$ to an adjacent one with $S<0$ by only slightly tuning the level position $\epsilon$.

The physical origin of the peaks are new channels of transport in which energy quanta absorbed from the environment lift electrons from the depth of the Fermi sea to the Fermi edge, where they produce a strong thermoelectric response. In contrast to a classical drive (i.e., an environment in a coherent state) the number of absorbable quanta and, hence, the number of Mott-features at negative energies is strictly limited by the prepared Fock-state.

\begin{figure}[t]
	\centering
	\includegraphics[width=8cm]{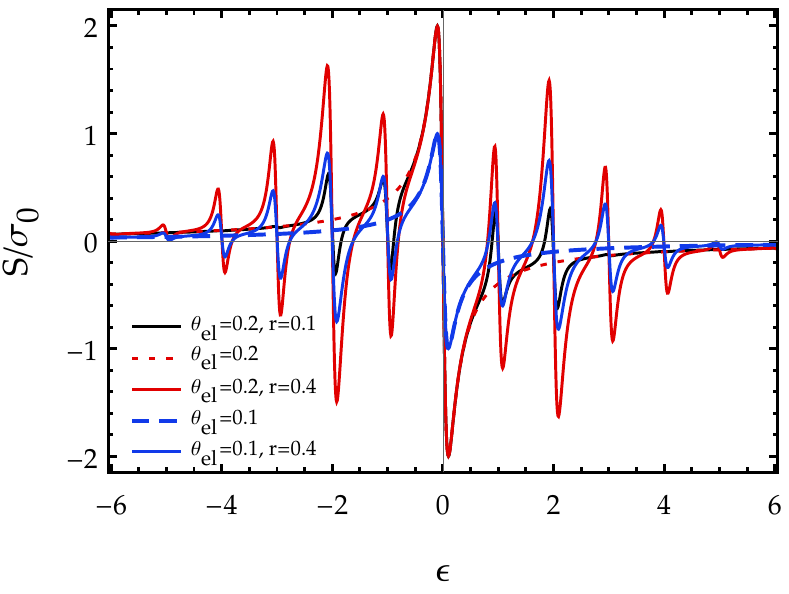}
	\caption{(Color online)
		Thermopower $S_0^{\mathrm{inel}}$ for a cavity in a squeezed state ($r \neq 0$). A squeezed state provides enhanced probabilities for emission and absorption of energy over a range increasing with the strength of squeezing. Comparing the amplitude of the peaks at energies $\epsilon=\pm 1$ with peaks at $\epsilon=\pm 2$ an indication of the even-odd asymmetry present in a squeezed state with real-valued squeezing parameter becomes visible. Units are scaled with $\hbar\Omega$ and $\gamma=0.1$. Results for $\rho=1$ (solid) and ballistic ones for $\rho=0$ (dashed) are shown.
	}
	\label{fig_S-ST-squeezed_T-dep}
\end{figure}

\subsection{Squeezed-state cavity}
A squeezed state in a cavity can be realized via reservoir engineering or driving of the cavity with squeezed vacuum noise. For such an  environment the number of quanta, which can be exchanged between cavity and junction, is not limited, since the cavity state includes arbitrary large excitations. With increasing photon number the probabilistic weight for emission is reduced though. To derive the corresponding distribution $P_{\mathrm{occ}}({E})$, one starts from its Fourier transform in time which reads
\begin{align}
\tilde{P}_{\mathrm{occ}}(\tau) = & \exp\left[ -4 \rho \sinh^2(r) \sin^2 \left( \frac{\Omega\tau}{2} \right) \right]  \nonumber \\
& \times\, I_0\left[ 2 \rho \sinh (2r) \sin^2 \left( \frac{\Omega\tau}{2} \right) \right]
\end{align}
with squeezing parameter $r$ and a modified Bessel function $I_0$ of the first kind. The total distribution $P_{\mathrm{tot}}(E)$ then follows according to (\ref{eq_P_tot}) as the Fourier transform of
\begin{equation}
\tilde{P}_{\mathrm{tot}}(\tau)=\tilde{P}_0(\tau)\cdot \tilde{P}_{\mathrm{occ}}(\tau).
\end{equation}
Since $\tilde{P}_{\mathrm{occ}}(\tau)$ is periodic in time with period ${2\pi}/{\Omega}$, it is sufficient to calculate the  Fourier-coefficients
\begin{equation}
p_{\mathrm{tot}}[k] = \frac{\Omega}{2\pi} \int\limits_0^{\frac{2\pi}{\Omega}} \mathrm{d}\tau \, \tilde{P}_{\mathrm{tot}}(\tau)\,  {\rm e}^{ -i k \Omega \tau}
\end{equation}
which determine $P_{\mathrm{tot}}(E)$ according to (\ref{eq_P_tot}).

\begin{widetext}
	
	\begin{figure}[h!]
		\centering
		\includegraphics[width=17.5cm]{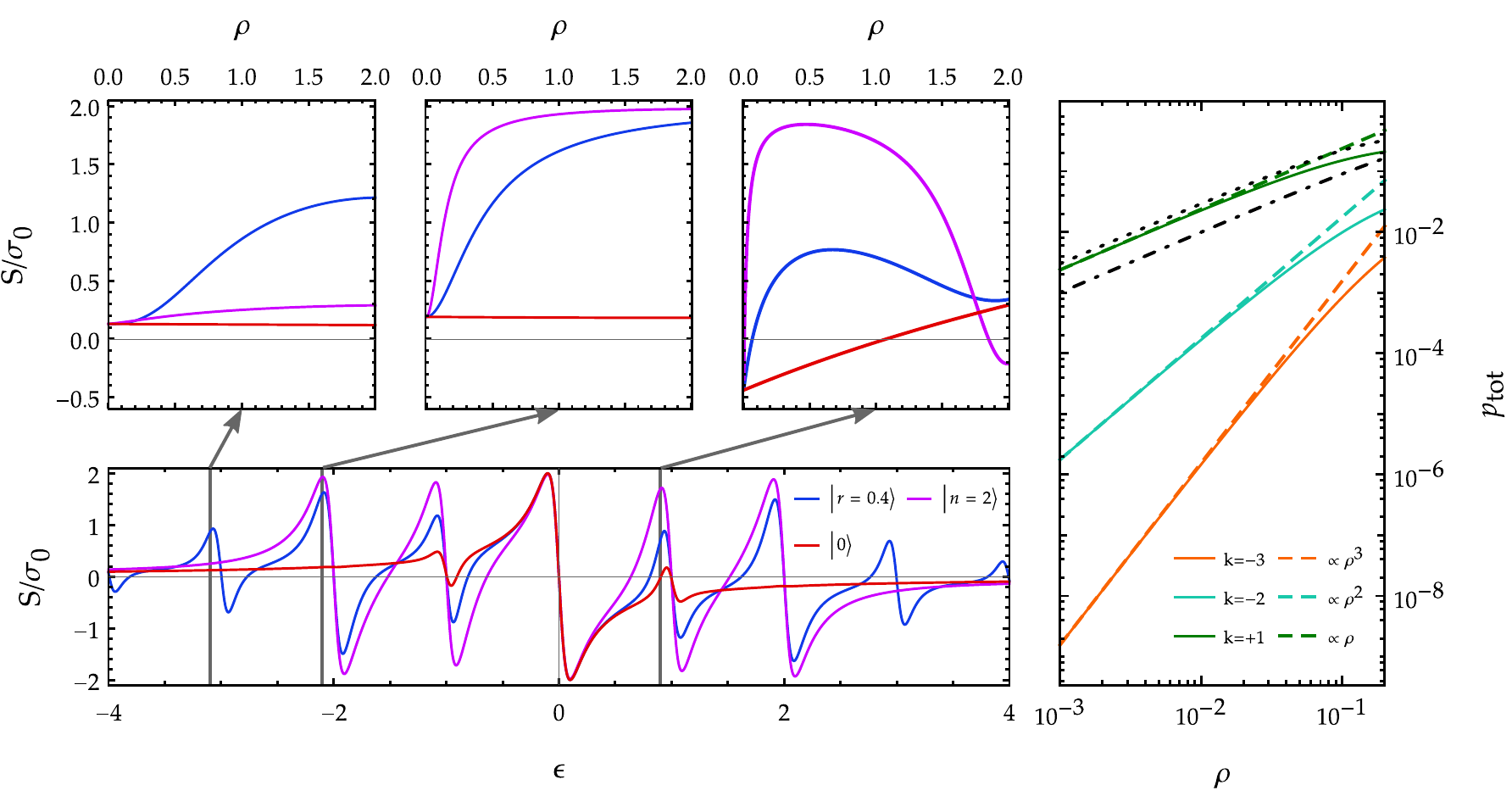}
		\caption{(Color online)
			Left: Comparison of the thermopower for the three considered environments: empty cavity (red), Fock state (magenta), and  squeezed state (blue) at $\theta_{\mathrm{el}}=0.2$, $\rho=1$, $\gamma=0.1$. The upper panels show the coupling strength dependence of $S(\epsilon = k\Omega -\gamma)$ for $k=-3$, $k=-2$, and $k=+1$ respectively. For weak coupling $\rho$ a generic scaling with $\rho^{|k|}$ is observed for $k$-photon emission and absorption for all environments. It can be traced back to a corresponding scaling of $p_{\mathrm{tot}}[k]$. Units are scaled with $\hbar\Omega$.\\
			Right: Weights $p_{\mathrm{tot}}[k]$ entering the energy-exchange function $P_{\mathrm{tot}}(E)$ for a squeezed state cavity, $|r=1\rangle$. Fock state (dots) and empty cavity (dot-dash) show the same generic scaling (shown for $k=1$).
		}
		\label{fig:rho-dependence}
	\end{figure}	
	
\end{widetext}

The thermopower versus the level position $\epsilon$ for various electronic temperatures $\theta_{\mathrm{el}}$ and squeezing parameters $r$ is shown in Fig.~\ref{fig_S-ST-squeezed_T-dep} together with the corresponding ballistic results ($\rho=0$).
Significant differences of Fig.~\ref{fig_S-ST-squeezed_T-dep} to the previous cases can be traced back to two important characteristics of a squeezed state, namely even-odd asymmetry and contributions with a large number of excitations. The latter is reflected in the appearance of many peaks in the thermopower for sufficiently strong squeezing and coupling (cf. Sec.~\ref{sec:rho-dependence} below). Only weak remnants of the strong even-odd asymmetry in the cavity occupation are inherited by the energy-exchange function, $P_{\mathrm{tot}}(E)$, and become visible in the relative height of the double-peak structure in the thermopower. In Fig.~\ref{fig_S-ST-squeezed_T-dep} (cf. also Fig.~\ref{fig:rho-dependence}) a suppression of the odd-absorption peak at $\epsilon = \pm 1$ as compared to the (even) two-quanta absorption at $\epsilon = \pm 2$ is observable for strong coupling $\rho=1$. 

\subsection{Coupling-strength dependence \label{sec:rho-dependence}}
In the following, we study in more detail, how the strength of the coupling $\rho$ to the environment influences the thermopower features connected to the various inelastic transport channels. For that purpose, Fig.~\ref{fig:rho-dependence} compares for all three investigated environments \-- empty cavity, Fock state $|n_0=2\rangle$, and  squeezed state $|r=0.4\rangle$ \-- how the heights of the thermopower peaks associated with the absorption of two and three quanta and the emission of a single cavity excitation depend on the coupling strength.

In general, thermopower features depend in a nontrivial manner on the strength of the underlying transport process, due to the fact, that there is always a ratio (namely of charge and energy transport) involved. For weak coupling $\rho$, however, the inelastic contribution to the thermopower around the channel $\epsilon = k \Omega$ scales with coupling strength in the same manner  as the corresponding energy-exchange function $P_{\mathrm{tot}}(E = k \Omega) \propto p_{\mathrm{tot}}[k]$ namely with  $\rho^{|k|}$. The reason being, that the same $\rho^{|k|}$ factor for $k$-photon emission and absorption enters the contribution in numerator and denominator via $p_{\mathrm{tot}}[k]$. If changes in the non-resonant background are negligible, an expansion for small coupling strength $\rho$  yields the same scaling for the (inelastic contribution to the) thermopower, $S(\rho)-S(\rho=0)$.

The right panel of Fig.~\ref{fig:rho-dependence} shows the scaling of the underlying weights $p_{\mathrm{tot}}[k]$ for a squeezed state and various photon-emission/absorption channels and comparisons to the case of empty cavity and Fock-state. The same scaling is observed in the linear/quadratic/cubic behavior of $S(\epsilon= k -\gamma)$ in the left panel of Fig.~\ref{fig:rho-dependence}.

For larger coupling strength $\rho$, the behavior is no longer generically determined by the number of exchanged quanta, but depends on the specific state of the environment, as reflected in the complex (oscillating) $\rho$-dependence, cf. for instance Eq.~\eqref{eq:p_tot_Fock} and Fig.~\ref{fig:rho-dependence}. For the squeezed state, one can again observe remnants of the even-odd asymmetry in the fact, that the one-photon contribution is reduced compared to the the two-photon contribution for sufficiently large coupling.

\section{Conclusion}
\label{sec:conclusion}
The goal of this work was to analyze the thermopower of tunnel junctions in contact with thermal and steady state reservoirs.  The junction itself consists of a resonant level strongly asymmetrically coupled to Fermi liquids in the leads which allows for transparent expressions and a detailed physical understanding. As a main result, we find that the thermopower can be controlled over a wide range including negative and positive values by simply tuning the level position $\epsilon$ with respect to the Fermi level or by changing properties of the environment.

For a purely ohmic impedance we observe a non-vanishing thermopower even for zero electronic and environmental temperature due to the ability of the environment to absorb energy. This is a significant improvement compared to purely ballistic transport, in which case the thermopower would vanish for the given limit of zero electronic temperature. Non-classical environments give rise to detailed resonance-like patterns in the thermopower associated with the exchange of single or multiple photon quanta between transferred charges and reservoirs. The ratio of the charging energy to the excitation quantum in the reservoir then defines the coupling constant. In principle it could be tuned in situ (for example by varying the inductance of an $LC$-circuit resonator) which provides another knob to vary the thermopower. Eventually, at low electronic temperatures the thermopower is an easy measure to obtain information about relevant transfer channels and the charge-reservoir correlation. Our findings thus provide not only an elegant way to describe the thermopower in actual circuits as long as the charge transfer is predominantly sequently, but may also be used to optimize the figure of merit $ZT$ of thermolelectric devices which depends quadratically on the thermopower.

\section*{Acknowledgements}
We thank Mohammad Mehmandoost for interesting discussions and acknowledge financial support through the IQ$^{\rm ST}$ and from the German Science Foundation (DFG) through SFB/TRR21 and AN 336/6-1.

\appendix
\bibliography{references.bib}

\end{document}